\documentclass[aps,superscriptaddress,twocolumn,a4paper,pra]{revtex4}
\usepackage{latexsym,graphicx,amsmath,amsfonts}

\def\ket#1{| #1 \rangle}

\def\bra#1{\langle #1 |}

\def\proj#1{\ket{#1}\bra{#1}}

\newcommand{\one}{\mathbb{I}}

\newcommand{\ee}{\end{equation}}
\newcommand{\bea}{\begin{eqnarray}}
\newcommand{\eea}{\end{eqnarray}}

\begin{document}
\title{Measuring Nothing}
\author{Daniel~K.~L.~\surname{Oi}} \email{daniel.oi@strath.ac.uk}
\affiliation{SUPA Department of Physics, University of Strathclyde,
  Glasgow G4 0NG, United Kingdom} \date{\today}

\author{V\'aclav~\surname{Poto\v cek}} 
\affiliation{Czech Technical University in Prague,
Faculty of Nuclear Sciences and Physical Engineering, Department of Physics,
B\v rehov\'a 7, 115 19 Praha 1, Czech Republic}

\author{John~\surname{Jeffers}} 
\affiliation{SUPA Department of Physics, University of Strathclyde,
  Glasgow G4 0NG, United Kingdom}

\begin{abstract}
  Measurement is integral to quantum information processing and
  communication; it is how information encoded in the state of a
  system is transformed into classical signals for further use. In
  quantum optics, measurements are typically destructive, so that the
  state is not available afterwards for further steps. Here we show
  how to measure the presence or absence of the vacuum in a quantum
  optical field without destroying the state, implementing the ideal
  projections onto the respective subspaces. This not only enables
  sequential measurements, useful for quantum communication, but it
  can also be adapted to create novel states of light via bare raising
  and lowering operators.
\end{abstract}

\maketitle

\section{Introduction}

At first glance, measuring the vacuum is trivial, a perfect
photodetector will reveal the vacuum state upon the non-occurrence of
a click. However, the converse result, i.e. not measuring the vacuum,
ideally should preserve the information in the non-vacuum sector for
further interrogation - something which is difficult to achieve with
direct photodetection.  Formally, we would like to implement the
following measurement projectors, $\{\proj{0},\one-\proj{0}\}$, where
the latter non-vacuum outcome removes the vacuum component without
affecting the relative amplitudes or coherences of the other Fock
states. This is crucial for sequential measurement schemes, such as
in~\cite{AnderssonOi2008,GTW2012,Sen2011}, and rules out other
projective schemes such as quantum nondemolition measurements of
photon number~\cite{Grangier1998,Guerlin2007}. The development of
practical methods for non-destructive measurements on optical
fields~\cite{Paris2002,Kok2002} is therefore an important topic for
future practical quantum information processing systems.

Measurement is also a key element in performing non-Gaussian
operations, e.g. for entanglement purification of continuous variable
states. Recent examples include the implementation of the quantum
optical creation and annihilation operators, both of which rely on
postselection \cite{par+-1,zav+-2,fiu+-3}. Extending the type of
possible operations is crucial for the production of tailored states
in quantum information systems. Our method can be simply extended to
provide a first realization of the bare photon addition and
subtraction operators. 

\begin{figure}
  \includegraphics[width=\columnwidth]{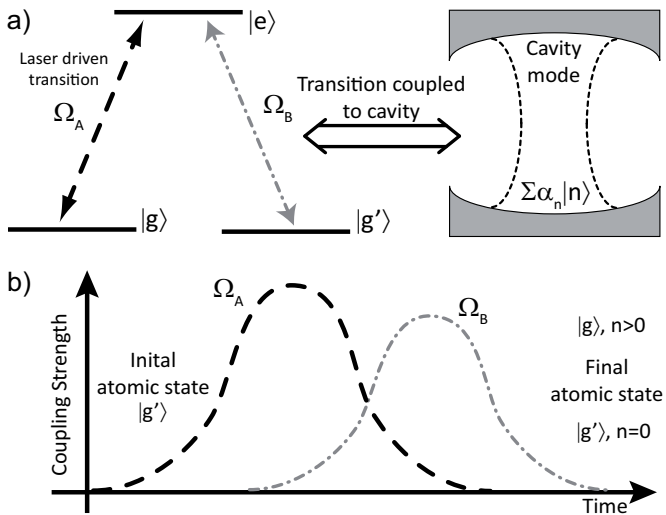}
  \caption{a) Lambda atomic system coupled to cavity. A three level
    atom with two ground states ($\ket{g}$, $\ket{g'}$) and a single
    excited level ($\ket{e}$) can be used to probe for the vacuum
    component of an optical field with annihilation operator $a$. The
    A transition is driven by a STIRAP laser and the B transition can
    be controllably coupled to the cavity mode to be
    measured. Initially, the atom is in the state $\ket{g'}$. The
    final state of the atom $\ket{g}$ or $\ket{g'}$ depends on the
    presence or absence of photons in the field respectively. b)
    Counter-intuitive pulse sequence for the couplings. Consider the
    case where there are $n\ge 1$ photons in the field. With the
    initial state of the combined atom-mode system $\ket{g',n}$,
    coupling A is turned on first. As coupling B is slowly increased,
    the atom-cavity state adiabatically follows the dark-state manifold
    $\sin\theta\ket{g,n-1}-\cos\theta\ket{g',n}$,
    $\theta=0\rightarrow\pi/2$. Coupling A is now turned off, followed
    by coupling B, leaving the final state of the system as
    $\ket{g,n-1}$. If the cavity was originally vacuum ($n=0$), the
    final state of the atom remains as $\ket{g'}$.}
\label{fig:lambdaatom}
\end{figure}

We consider a single mode of an optical cavity in an arbitrary quantum
state, $\rho$, as our system to be measured. To perform the
measurement, we introduce a probe which consists of a three level atom
in the $\Lambda$-configuration (See Fig.~\ref{fig:lambdaatom}a). The
cavity mode can be controllably coupled to transition B whereas
transition A interacts with an externally applied laser
field~\cite{PMZK1993,PMZCK1995}. In these papers the general adiabatic 
mapping of atomic levels to cavities was introduced. Our particularly 
simple configuration is insensitive to all field amplitudes other than 
the vacuum.

The Hamiltonian of the combined system can be written in the rotating
wave approximation as
\begin{equation}
\label{rwa}
\begin{aligned}
H^\mathit{RWA} &= \hbar \Delta |e\rangle\langle e| + \hbar\gamma_A(t) 
(|e\rangle\langle g| + |g\rangle\langle e|) \\ &\quad+ \hbar\gamma_B(t) 
(|e\rangle\langle g'| a + |g'\rangle\langle e| a^\dag),
\end{aligned}
\end{equation}
where the coupling constants $\gamma_A$ and $\gamma_B$ between the
atom and the two fields depend on the strength of the respective
fields at the point where the atom is located. An optional detuning
$\Delta$ can be applied to both fields in order to suppress
single-photon resonance effects as long as we maintain the two-photon
resonance condition,
\begin{equation}
E_g' - E_g = \hbar(\omega_B - \omega_A).
\end{equation}
The situation is similar to the V-STIRAP scheme for producing single
photons~\cite{Kuhn1999} where a cavity evolves from
$\ket{0}\rightarrow\ket{1}$ through a dark state adiabatic evolution
of an atom $\ket{g}\rightarrow\ket{g'}$.

In our measurement procedure, we run the V-STIRAP sequence in reverse:
the initial state of the atom is $\ket{g'}$, and the order of the A
and B couplings is switched (see Fig.~\ref{fig:lambdaatom}b). If the
cavity field initially contains at least one photon, at the end of the
sequence the atom is left in $\ket{g}$ and the field has one photon
subtracted. However, if the cavity was originally in the vacuum state,
the atom stays in $\ket{g'}$ and the cavity is left unchanged. An
initial superposition of the cavity evolves as
\begin{equation}
\ket{g'}\sum_{n=0}^{\infty}\alpha_n \ket{n}\rightarrow\ket{g'}\alpha_0\ket{0}
-\ket{g}\sum_{n=1}^\infty \alpha_n \ket{n-1}.
\end{equation}
The state of the atom is now entangled with that of the cavity. By
measuring the atomic state in either $\ket{g}$ or $\ket{g'}$, we have
determined whether the initial cavity state had at least one photon or
none. By coherent rotations of the ground states before a population
measurement, projections onto more general subspaces are also
possible.

If the atom is found in $\ket{g}$, the field amplitudes have been
shifted by one. The ideal projection $(\one-\proj{0})$ results if we
replace the subtracted photon, this is simply achieved by running the
V-STIRAP procedure forwards. Note that this does not require the
initial cavity state to be vacuum, we can add a photon to an arbitrary
state of the field. As discussed later, the shifting property of the
procedure can be exploited to perform novel operations and generate
non-classical states of light.

The key aspect of the adiabatic process is that the evolution of the
system does not rely on the dynamics of the Hamiltonian, provided that
the conditions of adiabatic transition are satisfied. In this way, the
state of the ancilla atom can be made asymptotically insensitive to the cavity
photon number, except for the critical case of the vacuum. In the
usual Jaynes-Cummings scenario, the dynamics in each of the combined
Fock subspaces proceeds at a rate proportional to the square root of
photon number, leading in general to different states of the atom. In
our scheme, the atom does not distinguish between different photon
numbers $n=1,2,3,\ldots$, which allows us to perform the ideal
projection onto the complement of the vacuum, in contrast to
previous proposals for quantum non-demolition measurements of the
optical field~\cite{Guerlin2007,Grangier1998}.

We can extend the method to project onto the joint $n$-mode vacuum
state or complement, as required in the decoding scheme
of~\cite{GTW2012}. This requires a probe atom with $n+2$ levels in an
$(n+1)$-pod configuration. Let $\ket{g_0}$ denote the initial state
of the atom, the remaining ground states be denoted $\ket{g_n}$ for
$n=1,\ldots,n$, and $\ket{e}$ be the excited level. The
$\ket{e}-\ket{g_n}$ ($n>0$) transitions are driven by lasers with
strength $\Gamma_j$ and the $\ket{e}-\ket{g_0}$ transition is
selectively coupled to each of the $n$-modes in turn with strength
$\gamma_j$. In real atoms, this may be difficult but it may be simpler
in engineered systems, e.g. superconducting qudits coupled to
transmission lines. We apply in turn the same procedure as for the
single mode measurement by sequential pairwise adiabatic variation of
$\{\Gamma_j,\gamma_j\}$, $j=1,\ldots,n$, after which the population of
$\ket{g_0}$ is determined. If the atom is detected in $\ket{g_0}$,
then the $n$-modes are projected onto the joint vacuum state
$\ket{00\dots 0}$, otherwise the atom and $n$-modes are left in a
(generally entangled) state where the $\ket{g_0}\ket{00\ldots0}$ state
has been truncated. To disentangle the atom and add back subtracted
photons, running the sequence of couplings backwards and in reverse
order returns the atom to $\ket{g_0}$ which erases any
information of the photon number distribution of the $n$-modes.

\section{Implementation}

\begin{figure}
\includegraphics[width=\columnwidth]{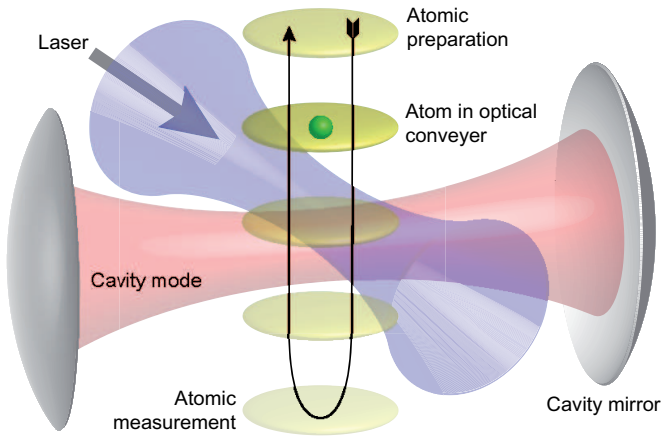}
\caption{Cavity QED vacuum measurement. The configuration of the
  cavity mode, driving laser and trajectory of the atom are similar to
  the V-STIRAP scheme, except that the laser beam is encountered
  before the cavity mode. An optical lattice traps and controls the
  position, hence coupling, of the atom with both the driving laser
  and cavity mode. State preparation in $\ket{g'}$ is performed before
  the atom is transported into the cavity. After the atom has crossed
  the cavity and adiabatically interacted with laser and mode, it is
  measured, e.g. by fluorescence shelving~\cite{OxfordIon2008} or
  cavity enhanced detection~\cite{TONJFO2009}, to
  discover which ground state it is in. To perform the ideal
  $(\one-\proj{0})$ operation in the case of the $\ket{g}$ result, the
  motion of the atom is reversed in order to replace the photon
  extracted from the cavity.}
\label{fig:experiment}
\end{figure}

\begin{figure}
\includegraphics[width=\columnwidth]{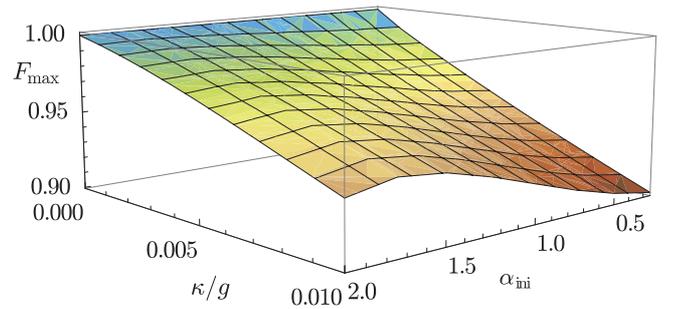}
\caption{Fidelity of the full measurement procedure, conditioned on
  the detection of the atom in $|g\rangle$. The initial cavity field
  is in a coherent state of amplitude $\alpha_{ini}$. The coupling constants
  $\gamma_A$ and $\gamma_B$ were modulated as $\cos^2t$ and $\sin^2t$,
  respectively, and their amplitudes were in a $2 : 1$ ratio. The
  detuning $\Delta$ was zero and the decay rate of the $|e\rangle$
  state was $1\%$ of the maximum cavity coupling $g = \max
  \gamma_B$. The cavity loss parameter $\kappa$ is also plotted
  relative to $g$. Assuming the knowledge of the cavity properties,
  the fidelity optimizing time of transition was chosen in each data
  point. We note that assuming an educated guess for the initial mean
  photon number of the state, the point of maximum fidelity at a
  constant loss may be shifted along the $\alpha$ axis by altering the
  intensity of the laser field.}
\label{fig:decoh}
\end{figure}

The experimental setup for V-STIRAP~\cite{Kuhn1999} can be adapted to
perform our vacuum measurement (Fig.\ref{fig:experiment}). We use a
cavity with a long storage time to reduce leakage and decoherence. We
also introduce preparation and readout zones for the atom. The motion
of the atom is reversed in the case of measuring the atom in $\ket{g}$
in order to replace a subtracted photon. There are several
experimental challenges, mainly the lifetime of the field compared to
the time required to implement the measurement. The cavity field must
last long enough for the atom to be adiabatically transported,
measured, and returned. The damping rate of the cavity and atom-cavity
coupling are both highly dependent upon the effective mode volume of
the cavity and balancing these factors will be system dependent. To
give an indication of the performance of the protocol under non-ideal
condition, we have simulated the measurement of a lossy cavity
with finite sweep times, the results displayed in
Fig.\ref{fig:decoh}.

\section{Applications}

A straightforward application of this measurement is in sequential
decoder schemes as discussed in~\cite{Sen2011}. In the protocol
of~\cite{GTW2012}, the state of a $n$-mode system has to be
identified. The state is taken from an ensemble of products of
coherent states,
$\{\ket{\alpha_1^k,\alpha_2^k,\ldots,\alpha_n^k}\}$. A sequence of
displacements and projections onto the $n$-mode vacuum or its
complement has been shown to decode the message successfully in the
$n\rightarrow\infty$ limit as long as the rate of transmission is
below the Holevo bound.

We can also use the photon number altering properties of our procedure
to enact bare raising and lowering operations, in contrast to the
creation $a^\dagger$ and annihilation operators $a$ as usually
considered. The non-Hermitian $a^\dagger$ and $a$ operators represent
non-Gaussian operations and have been realized probabilistically in
experiments~\cite{par+-1,zav+-2,fiu+-3}.  ``Subtracting'' a photon
from squeezed light can produce an approximate Schr\"odinger cat
state~\cite{cattheory,cat1,cat2,cat3}, and both processes have been
used in super-optimal optical amplification
protocols~\cite{marek,fiurasek,zavatta,usugua}.

However, the $a^\dagger$ and $a$ operators do not simply add and subtract 
photons, but also Bose condition the state. Pure addition and 
subtraction of photons are represented by bare 
raising and lowering operators~\cite{susskglog} (sometimes known as 
photon number shifting operators~\cite{lee})
\begin{equation}
E^+= \sum_{n=0}^\infty |n+1 \rangle \langle n|,\quad
E^-= \sum_{n=1}^\infty |n-1 \rangle \langle n|.
\end{equation}
These can produce nonclassical states of light; for example any state
which has $E^+$ applied to it must violate the Klyshko
criterion~\cite{lee}. Applying $E^+$ to a coherent state
produces a state with subpoissonian statistics, whereas applying
$\hat{E}^-$ makes it superpoissonian.

There has been little study of the bare operators and their effects,
mainly because they have not been realized
experimentally~\cite{benaryeh}. Implementing $E^+$ and
$E^- $ requires cancellation of the $\sqrt{n}$ Bose enhancement
factors inherent in $\hat{a}$ and $\hat{a}^\dagger$. The nature of the
$(\one-\proj{0})$ projection and the adiabatic process that we have
described does not alter the relative weights of the amplitudes
corresponding to different photon numbers, in contrast to other
schemes which rely on $a^\dagger$. The V-STIRAP process implements 
$E^+$ and the reverse process realises $E^-$. In addressing the problem of
quantum optical phase the measurement of moments of bare operators was
proposed using a basic scheme similar to that considered here, without
a detailed analysis of the effect of reachable experimental
parameters~\cite{ZSG2006}. We will explore the detailed consequences
of this elsewhere.

\begin{figure}
\includegraphics[width=\columnwidth]{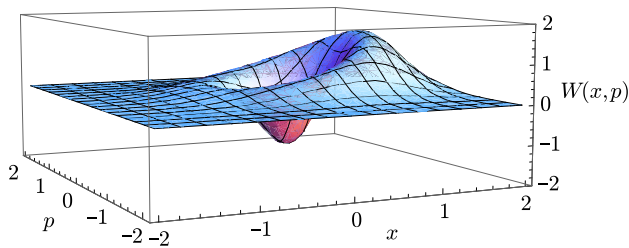}
\caption{Wigner function of a vacuum-stripped $\ket{\alpha=1}$
  coherent state. In the protocol a loss rate of
  $\kappa=0.005 g$ and other parameters as specified in
  Fig.~\ref{fig:decoh} were used. It shares all the features of the
  idealised output state, including negativity. This state is obtained
  from $\ket{\alpha=1}$ with a probability of $0.61$. The probability
  of measuring the vacuum $|\langle 1|0 \rangle|^2$ is $0.37$, and a
  $2\%$ error is caused by photon loss during the process. The
  fidelity is $96\%$ compared with $(\one-\proj{0}) \ket{\alpha=1}$.}
\label{fig:wigner}
\end{figure}

We can further exploit the conditional dynamics preserving the
relative amplitudes for all non-zero number states to implement a
reverse quantum scissors~\cite{Pegg1998}. In the original quantum
scissors scheme a general superposition of photon numbers has photon
numbers higher than 1 removed, without altering the zero and one
photon amplitudes. The scheme has been theoretically extended to make
the cut at higher photon numbers \cite{KKGJ2000,J2010}. By successive
application of the measurement without photon replacement $n$ times,
we can truncate the first $n$ amplitudes of a state, conditioned on
not observing the vacuum. By adding $n$ photons, we return the state
to its original form but without the first $n$ terms
(Fig.\ref{fig:wigner}). The probability that this will occur is
$1-\sum_{k=0}^{n-1}P_k$ where $P_k$ is the probability of observing
$k$ photons. Similarly, we also can simply use the protocol to perform
photon number resolving measurements.

\section{Conclusion}

The ability to implement ideal projections on a field opens up new
possibilities for quantum communication and computation. Adiabatic
evolution in our method avoids the $\sqrt{n}$ factors in dynamical
schemes and achieves the unusual nonlinearity required. Though it will
be challenging to engineer systems with the requisite long storage
times with strong coupling, recent advances in microcavities as well
as microwave and nanomechanical systems give grounds for optimism. The
simplicity and utility of the system described here for implementing
several quantum optical information protocols should be significant
drivers towards this goal.

\appendix

\section{Technical Details}

To see why the methods works, we examine the form of the rotating-wave
Hamiltonian for the atom-field system,
\begin{equation}
\begin{aligned}
H^{\sl RWA} &= \hbar \Delta |e\rangle\langle e| + \hbar\gamma_A(t) 
(|e\rangle\langle g| + |g\rangle\langle e|) \\ &\quad+ \hbar\gamma_B(t) 
(|e\rangle\langle g'| a + |g'\rangle\langle e| a^\dag),
\end{aligned}
\end{equation}
Firstly, we note that for any $n \in \mathbb{N}$, the subspace
$\mathop\{|g,n-1\rangle, |e,n-1\rangle, |g',n\rangle\}$ is coupled
together by $H^{\it RWA}$. Other states are not coupled to this
triplet, and this simplifies the dynamics significantly.  Together
with the linear span of the vector $|g',0\rangle$, which itself is an
eigenspace corresponding to the eigenvalue $0$, these subspaces allow
for a decomposition of the whole Hilbert space describing the coupled
system of the atom and the cavity. Within each of the
three-dimensional subspaces, we can identify three nondegenerate
eigenenergies: $0$ and $\frac{\hbar}{2} (\Delta \pm \sqrt{\Delta^2 +
  4\gamma_A^2(t) + 4n \gamma_B^2(t)})$. The eigenstates corresponding
to zero energy are, up to normalisation and phase factors, $\sqrt{n}
\gamma_B(t) |g,n-1\rangle - \gamma_A(t) |g',n\rangle$, along with the
special case of $|g',0\rangle$, noted above.

In order to study the deviations from an ideal adiabatic transition, we solve 
the time-dependent Schr\"odinger equation. For a given $n$, we 
denote
\begin{equation}
\theta(t) = \arctan\frac{\sqrt{n}\gamma_B(t)}{\gamma_A(t)}
\end{equation}
and
\begin{equation}
\nu(t) = \sqrt{\gamma_A^2(t) + n\gamma_B^2(t)}.
\end{equation}
In this notation, the dark state can be expressed as
\begin{equation}
|a(t)\rangle = \sin\theta(t) |g,n-1\rangle - \cos\theta(t) |g',n\rangle.
\end{equation}
We rewrite the equation of motion in an orthonormal basis consisting of the 
vectors $|a(t)\rangle$,
\begin{equation}
|b(t)\rangle = \cos\theta(t) |g,n-1\rangle + \sin\theta(t) |g',n\rangle,
\end{equation}
and $|e\rangle$.

Resolving the instantaneous state as
\begin{equation}
|\psi(t)\rangle = \alpha_a(t) |a(t)\rangle + \alpha_b(t) |b(t)\rangle 
+ \alpha_e(t) |e\rangle,
\end{equation}
we substitute this into the Schr\"odinger equation to find the
equations of motion for $\alpha_i(t)$,
\begin{equation}
\begin{aligned}
\dot\alpha_a &= \dot\theta \alpha_b, \\
\dot\alpha_b &= -\dot\theta \alpha_a - i\nu \alpha_e, \\
\dot\alpha_e &= -i\nu\alpha_b - i\Delta \alpha_e.
\end{aligned}
\end{equation}
If the system begins in the dark state the initial conditions are $\alpha_a(0)=1$, $\alpha_b(0) = \alpha_e(0) = 0$. The equations are best solved in terms of the projective space coordinates 
$\kappa_b = \alpha_b/\alpha_a$ and $\kappa_e = \alpha_e/\alpha_a$, where they 
become
\begin{equation}
\begin{aligned}
\dot\kappa_b &= -\dot\theta - i\nu\kappa_e - \dot\theta\kappa_b^2, \\
\dot\kappa_e &= -i\nu\kappa_b - i\Delta\kappa_e - \dot\theta\kappa_b\kappa_e.
\end{aligned}
\end{equation}

The last pair of equations can be solved asymptotically. For this
purpose, we denote the complete time of transition $T$, such that all
of $\dot\theta$, $\ddot\theta/\dot\theta$ and $\dot\nu/\nu$ are upper
bounded by a constant multiple of $T^{-1}$. Also, we denote $\nu_0$
the minimum modulus of the lower of the nonzero eigenvalues of $H^{\it
  RWA}$ (restricted to the given subspace) reached during the
transition. This value is also a lower bound for both of $\nu(t)$ and
$\nu^2(t)/\Delta$. We use these inequalities to find that
\begin{equation}
\label{eq-kappa}
\begin{aligned}
\kappa_b(t) &= -i\frac{\dot\theta(t) \Delta}{\nu^2(t)} + O((\nu_0 T)^{-2}), \\
\kappa_e(t) &= i\frac{\dot\theta(t)}{\nu(t)} + O((\nu_0 T)^{-2}).
\end{aligned}
\end{equation}
These formulas allow us to express
\begin{equation}
\alpha_a(t) = \exp\left( -i\Delta \int_0^t \frac{\dot\theta^2(t)}{\nu^2(t)} 
\mathrm{d}t \right) + O((\nu_0 T)^{-2}).
\end{equation}

If the spatial distributions of intensity of the two light modes are smooth 
functions of coordinates, the time derivative of $\theta(t)$ for both $t=0$ and 
$T=0$ is zero. It immediately follows that the relative contributions of the 
states orthogonal to the desired final state, as given by Eq.~(\ref{eq-kappa}), 
vanish at the end of the transition, leaving only the asymptotic term. The 
probability of the diabatic transition is in turn $O((\nu_0 T)^{-4})$ and can be 
pushed arbitrarily close to zero by choosing a sufficiently long time $T$. The 
resulting state obtains a phase shift of
\begin{equation}
\phi = -\Delta \int_0^T \frac{\dot\theta^2(t)}{\nu^2(t)} \mathrm{d}t + O((\nu_0 
T)^{-2}),
\end{equation}
which itself scales as $O((\nu_0 T)^{-1})$ and also has a limit of zero for $T 
\rightarrow \infty$.

The dependence of the results on the photon number $n$ manifests itself through 
the constant
\begin{equation}
\nu_0 = \min_{t \in (0,T)} \left( \sqrt{\left(\frac\Delta2\right)^2 
+ \gamma_A^2(t) + n\gamma_B^2(t)} - \frac\Delta2 \right).
\end{equation}
This formula suggests that the scalings of both the probability of
error and the phase $\phi$ are actually the more favourable the higher
$n$ one operates in.  Studying the worst case, i.e., $n=1$, we find
that the transition is optimal if the two light modes overlap in such
a way that the effective beginning or end of either one coincides with
the point of maximal intensity of the other one.

We finish the analysis by noting that the state $|g',0\rangle$ remains
unchanged during the whole transition and obtains no phase shift due
to the fact that it corresponds to exactly zero energy.

As a result, any linear superposition
\begin{equation}
\sum_{n=0}^\infty \alpha_n |g,n\rangle
\end{equation}
evolves under ``forward'' STIRAP to
\begin{equation}
\sum_{n=0}^\infty \alpha_n |g',n+1\rangle,
\end{equation}
in a deterministic manner, up to correction terms of order $(\nu_0^{\it (min)} 
T)^{-1}$. Similarly, letting the atom encounter the laser field A first, any 
state of the form
\begin{equation}
\sum_{n=0}^\infty \alpha_n |g',n\rangle
\end{equation}
evolves into
\begin{equation}
\alpha_0 |g',0\rangle + \sum_{n=1}^\infty \alpha_n |g,n-1\rangle.
\end{equation}

In the case of a cavity in a mixed state, the corresponding operation is 
performed on each element of its Schmidt decomposition.

\section{Experimental Considerations}

For the implementation of the measurement, this requires that the
system being measured does not significantly evolve over the timescale
required to perform the adiabatic evolutions and atomic
measurement. This will be a considerable experimental challenge,
though advances in cavity QED have led to high coupling strengths
compared to loss rates which are the dominant sources of decoherence.

\subsection{Cavity lifetime}

Very high Q-factors have been achieved in optical resonators, of the
order of $3.5\times 10^{12}$ for large supermirror
cavities~\cite{Schreiber2012}, and approaching 10$^{11}$ for Fabry-Perot cavities~\cite{Kuhr2012}. In addition, the atom must couple
strongly with the cavity mode as this will limit the speed of the
atomic transport at which adiabaticity can be maintained. Together
with the low damping rate, we require that the atom-cavity system is
within the strong coupling regime. Large coupling will be favoured by
small mode volumes but this may conflict with the required storage
time. Ringdown measurements for QED experiments have shown a decay
time of 12.4 microseconds for a 1cm long cavity~\cite{Dotsenko2007}.

\subsubsection{Cavity-atom coupling}

The coupling $\gamma_B$ limits the rate at which the atom can traverse
the cavity. Coupling rates of the order of $10\,{\rm MHz}$ have been
achieved in cavity QED~\cite{Schuster2008}, which leads to an
adiabatic-safe interaction time on the order of $\sim 1$
microsecond. For a proof of principle experiment, the measurement of
the atom can take far longer than the decay time of the cavity if one
is not interested in restoring the subtracted photon. Much higher
couplings have been reported for atoms interacting with evanescent
field of a toroidal microresonator, of the order of $40\,{\rm
  MHz}$~\cite{ASALOVK2010}.

\subsubsection{Detection}

A standard method of state detection is shelving fluorescence. A long
cavity storage time is essential for the measurement as shelving
fluorescence takes a finite time, of the order of 100 microseconds for
99.9\% fidelity~\cite{OxfordIon2008} due to the need for the atom to
absorb and spontaneously emit photons and for these to be
detected. This time could be reduced by increasing the collection
solid angle and the probe power.

Faster atomic state detection could be achieved by using another
cavity~\cite{PBC2008}. If the atom is strongly coupled to this
detection cavity, this can affect the transmission of the cavity which
can be interrogated by a probe beam. In practice, a different mode of
the same cavity could be used for detection. A detection time of
10~microseconds at 97\% efficiency has been reported for fibre-coupled
cavity detection~\cite{GTKRH2011}. Even faster detection has been
achieved by combine both fluorescence and cavity coupling, 99.7\% in
less than 1 microsecond~\cite{TONJFO2009}. Detection of atoms coupled
to cavities in 250 nanoseconds has also been
reported~\cite{ASALOVK2010}.

\subsubsection{Other physical systems}

Photonic cavities may also be a possibility to tune the mode volumes,
coupling strengths and system geometry. Whispering gallery modes in
microspheres or microtoroids are also possibilities though the
required geometry is more complex. Controlling the atomic position and
shining the STIRAP laser may be issues. Projected
$\gamma_B=\mathrm{several}\ 100\,{\rm MHz}$, $\kappa < 1\,{\rm
  MHz}$~\cite{SpillanePRA} for such microtoroid cavities. Coupled
ion-cavity QED experiments may also be a viable system. The progress
on segmented traps and in shuttling~\cite{Amini2010} allows controlled
ion transport through a cavity field~\cite{Keller2007}.

We note that the same technique can be applied to
superconducting systems of artificial three-level
atoms~\cite{Bianchetti2010} and stripline resonators using adjustable
couplings~\cite{Bialczak2011,Gambetta2011}. In these systems, very
large couplings have been observed ($200\,{\rm MHz}$~\cite{DiCarlo2009}). It is
calculated that couplings of up to $360\,{\rm MHz}$ could be achieved in
optimized geometries~\cite{Bourassa2009}. Cavity loss of
$\kappa=250\,{\rm kHz}$ has been achieved in other
experiments~\cite{Schuster2007}. Tunable couplings between
superconducting elements have been developed
($(0-100)\,{\rm MHz}$)~\cite{Bialczak2011,Gambetta2011}. Measurement of three
level systems have been reported in~\cite{Bianchetti2010}.

\begin{acknowledgments}
\footnotesize

DKLO acknowledges discussions with Si-Hui Tan. The authors acknowledge
support from: Quantum Information Scotland (QUISCO), DKLO, JJ; the UK
EPSRC (EP/G009570/1), JJ, VP; Czech Science Foundation (GA\v CR
202/08/H072), VP.

\end{acknowledgments}


\begin{thebibliography}{99}
\bibitem{AnderssonOi2008}Andersson, E. \& Oi, D. K. L.
Binary search trees for generalized measurements.
\textit{Phys. Rev. A} \textbf{77}, 052104 (2008)

\bibitem{GTW2012}Guha, S., Tan, S-H., Wilde, M. M.
Explicit capacity-achieving receivers for optical communication and quantum
  reading. Pre-print at http://arXiv.org/abs/1202.0518 (2012)

\bibitem{Sen2011}Sen, P.
Achieving the Han-Kobayashi inner bound for the
quantum interference channel by sequential decoding.
Pre-print at http://arXiv.org/abs/1109.0802

\bibitem{Paris2002}Paris, M.
Optimized QND measurement of a field quadrature.
\textit{Phys. Rev. A} \textbf{65}, 012110 (2002)

\bibitem{Kok2002}Kok, P., Lee, H. \& Dowling, J. P.
Single-photon quantum nondemolition detectors constructed with linear optics and projective measurements.
\textit{Phys. Rev. A} \textbf{6}, 063814 (2002)

\bibitem{Guerlin2007}Guerlin, C. \textit{et al.} Progressive
  field-state collapse and quantum non-demolition photon counting.
  \textit{Nature} \textbf{448}, 889--893 (2007)

\bibitem{Grangier1998}Grangier, P., Levenson J. A. \& Poizat, J. P.
  Quantum non-demolition measurements in optics.
\textit{Nature} \textbf{396}, 537--542 (1998)

\bibitem{par+-1}Parigi, V., Zavatta, A., Kim, M. S. \& Bellini, M.
  Probing quantum commutation rules by addition and subtraction of
  single photons to/from a light field.
\textit{Science} {\bf 317}, 18901893 (2007).

\bibitem{zav+-2} Zavatta, A., Parigi, V., Kim, M. S., Jeong, H. \&
  Bellini, M. Experimental demonstration of the bosonic commutation
  relation via superpositions of quantum operations on thermal light
  fields. \textit{Phys. Rev. Lett.} {\bf 103}, 140406 (2009).

\bibitem{fiu+-3}Fiur\'a\v sek, J. Engineering quantum operations on
  traveling light beams by multiple photon addition and subtraction.
  \textit{Phys. Rev. A} {\bf 80}, 053822 (2009).

\bibitem{PMZK1993}Parkins, A. S., Marte, P., Zoller, P. \& Kimble,
  H. J.  Synthesis of arbitrary quantum states via adiabatic transfer
  of Zeeman coherence. \textit{Phys. Rev. Lett.} \textbf{71},
  3095--3098 (1993)

\bibitem{PMZCK1995}Parkins, A. S., Marte, P., Zoller, P., Carnal O. \&
  Kimble, H. J.  Quantum-state mapping between multilevel atoms and cavity light fields. \textit{Phys. Rev. A} \textbf{51}, 1578--1596 (1995)

\bibitem{Kuhn1999}Kuhn, A., Hennrich, M., Bondo, T. \& Rempe, G.
  Controlled generation of single photons from a strongly coupled
  atom-cavity system. \textit{Appl. Phys. B: Lasers and Optics} \textbf{69},
  373--377 (1999)


\bibitem{OxfordIon2008}
Myerson, A. H. \textit{et al.},
High-fidelity readout of trapped-ion qubits.
\textit{Phys. Rev. Lett.} \textbf{100}, 200502 (2008)

\bibitem{TONJFO2009}Terraciano, M. L. \textit{et
    al.}, 
  Photon burst detection of single atoms in an optical cavity.
  \textit{Nature Phys.} \textbf{5}, 480--484 (2009)
















\bibitem{cattheory} Dakna, M., Anhut, T., Opatrn\' y, T., Kn\"oll, L. \& Welsch, D.-G.
Generating Schr\"odinger-cat-like states by means of conditional measurements on a beam splitter.
\textit{Phys. Rev. A} \textbf{55}, 3184--3194 (1997).

\bibitem{cat1} Ourjoumtsev, A., Tualle-Brouri, R., Laurat, J. \& Grangier, P.
Generating Optical Schr\"odinger Kittens for Quantum Information Processing.
\textit{Science} \textbf{312}, 83--86 (2006).

\bibitem{cat2}Neergaard-Nielsen, J. S., Nielsen, B. M., Hettich, C., Molmer, K. \& Polzik, E. S.
Generation of a superposition of odd photon number states for quantum information networks.
\textit{Phys. Rev. Lett.} \textbf{97}, 083604(2006).

\bibitem{cat3}Wakui, K., Takahashi, H., Furusawa, A. \& Sasaki, M.
Photon subtracted squeezed states generated with periodically poled KTiOPO4.
\textit{Optics Express} \textbf{15}, 3568--3574 (2007).

\bibitem{marek}Marek, P. \& Filip, R.
Coherent-state phase concentration by quantum probabilistic amplification.
\textit{Phys. Rev. A} \textbf{81}, 022302 (2010). 

\bibitem{fiurasek}Fiur\'a\v sek, J.
Engineering quantum operations on traveling light beams by multiple photon addition and subtraction.
\textit{Phys. Rev. A} \textbf{80}, 053822 (2009). 

\bibitem{zavatta}Zavatta, A., Fiur\' a\v sek, J. \& Bellini, M.
A high-fidelity noiseless amplifier for quantum light states.
\textit{Nature Photonics} \textbf{5}, 52--60 (2011). 

\bibitem{usugua}Usuga, M. A. \textit{et al.}
Noise-Powered Probabilistic Concentration of Phase Information.
\textit{Nature Phys.} \textbf{6}, 767--771 (2010).

\bibitem{susskglog}Susskind, L. \& Glogower, J.
Quantum mechanical phase and time operator.
\textit{Physics} \textbf{1}, 49--61 (1964).

\bibitem{lee}Lee, C. T.
Application of Klyshko's criterion for nonclassical states to the micromaser pumped by ultracold atoms.
 \textit{Phys. Rev A} \textbf{55}, 4449--4453 (1997).

\bibitem{benaryeh}Ben-Aryeh, Y. \& Brif, C.
Discrete photodetection and Susskind-Glogower ladder operators.
Pre-print at http://arXiv.org/abs/quant-ph/9504009.

\bibitem{ZSG2006}Zou, X., Shu, J. \& Guo, G., Simple scheme for direct
  measurement of exponential phase moments of cavity fields by
  adiabatic passage. \textit{Physics Letters A} \textbf{359} 117--121
  (2006).

\bibitem{Pegg1998}Pegg, D. T., Phillips, L. S. \& Barnett, S. M.
Optical State Truncation by Projection Synthesis.
\textit{Phys. Rev. Lett.} \textbf{81}, 1604--1606 (1998)


\bibitem{KKGJ2000}Koniorczyk, M., Kurucz, Z., G\'abris, A. \& Janszky, J.
General optical state truncation and its teleportation.
\textit{Phys. Rev. A} \textbf{62}, 013802 (2000). 

\bibitem{J2010} Jeffers, J.
Optical amplifier-powered quantum optical amplification.
\textit{Phys. Rev. A} \textbf{82}, 063828 (2010).


\bibitem{Schreiber2012}K. U. Schreiber, K. U., Gebauer, A. \& Well, J.-P. R.
Long-term frequency stabilization of a $16 m^2$ ring laser gyroscope.
\textit{Opt. Lett.} \textbf{37}, 1925--1927 (2012)

\bibitem{Kuhr2012} S. Kuhr et al, Ultrahigh finesse Fabry-Perot superconducting resonator, 
\textit{Applied Physics Letters} \textbf{90}, 164101 (2007).

\bibitem{Dotsenko2007}
Dotsenko, I.
Single atoms on demand for cavity QED experiments.
PhD Thesis,
University of Bonn (2007)

\bibitem{Schuster2008}
  Schuster, I. \textit{et al.}, 
Nonlinear spectroscopy of photons bound to one atom.
\textit{Nature Physics} \textbf{4}, 382--385 (2008)

\bibitem{ASALOVK2010}
Alton, D. J. \textit{et al.} 
Strong interactions of single atoms and photons near a dielectric boundary.
\textit{Nature Phys.} \textbf{7}, 159--165 (2010)


\bibitem{PBC2008}
Poldy, R., Buchler, B. C. \& Close, J. D.
Single-atom detection with optical cavities.
\textit{Phys. Rev. A} \textbf{78}, 013640 (2008)

\bibitem{GTKRH2011}
Goldwin, J., Trupke, M., Kenner, J., Ratnapala, A. \& Hinds, E. A.
Fast cavity-enhanced atom detection with low noise and high fidelity.
\textit{Nature Comms.} \textbf{2}, 418 (2011)


\bibitem{SpillanePRA}
Spillane, S. M. \textit{et al.}
Ultrahigh-Q toroidal microresonators for cavity quantum electrodynamics.
\textit{Phys. Rev. A} \textbf{71}, 013817 (2005)

\bibitem{Amini2010}Amini, J. M. \textit{et al.}
Toward scalable ion traps for quantum information processing.
\textit{New J. Phys.} \textbf{12}, 033031 (2010)

\bibitem{Keller2007}Keller, K., Lange, B., Hayasaka, K., Lange, W. \& Walther, H.
Stable long-term coupling of a single ion to a cavity mode.
\textit{J. Mod. Opt.} \textbf{54}, 1607--1617 (2007)

\bibitem{Bianchetti2010}
Bianchetti, R. \textit{et al.}
Control and Tomography of a Three Level Superconducting Artificial Atom.
\textit{Phys. Rev. Lett.} \textbf{105}, 223601 (2010)

\bibitem{Bialczak2011}
Bialczak, R. C. \textit{et al.}
Fast Tunable Coupler for Superconducting Qubits.
\textit{Phys. Rev. Lett.} \textbf{106}, 060501 (2011)

\bibitem{Gambetta2011}
Gambetta, J. M., Houck, A. A. \and Blais, A.
Superconducting Qubit with Purcell Protection and Tunable Coupling, 
\textit{Phys. Rev. Lett.} \textbf{106}, 030502 (2011)

\bibitem{DiCarlo2009}
DiCarlo, L. \textit{et al.}
Demonstration of two-qubit algorithms with a superconducting quantum processor.
\textit{Nature} \textbf{460}, 240--244 (2009)

\bibitem{Bourassa2009}
Bourassa, J. \textit{et al.}
Ultrastrong coupling regime of cavity QED with phase-biased flux qubits.
\textit{Phys. Rev. A} \textbf{80}, 032109 (2009)

\bibitem{Schuster2007}
Schuster, D. I. \textit{et al.}
Resolving photon number states in a superconducting circuit.
\textit{Nature} \textbf{445}, 515--518 (2007)

\end{thebibliography}
\end{document}